\documentclass[pra,preprint,showpacs,letterpaper]{revtex4}
\usepackage{graphicx}
\usepackage{amsmath}
\usepackage{amsfonts}
\usepackage{bm}
\usepackage{color}

\begin{document}
\title{Backward scattering of low-energy antiprotons by highly charged 
and neutral uranium: Coulomb glory}
\date{\today}

\author{A.~V.~Maiorova}
\author{D.~A.~Telnov}
\author{V.~M.~Shabaev}
\author{I.~I.~Tupitsyn}
\affiliation{Department of Physics, St.~Petersburg State University, 
Ulianovskaya 1, Petrodvorets, St.~Petersburg 198504, Russia}
\author{G.~Plunien}
\affiliation{Institut f\"{u}r Theoretische Physik, TU Dresden, Mommsenstrasse
13, D-01062 Dresden, Germany}
\author{T.~St\"ohlker}
\affiliation{Gesellschaft f\"{u}r Schwerionenforschung, Planckstrasse 1,
D-64291 Darmstadt, Germany}
\begin{abstract}
Collisions of antiprotons with He-, Ne-, Ni-like, bare, and neutral uranium are studied theoretically for scattering angles close to 180$^{\circ}$ and antiproton energies 
with the interval 100 eV -- 10 keV. We investigate the Coulomb glory effect which is caused by a screening of the Coulomb potential of the nucleus and results in a prominent maximum of the differential cross section in the backward direction at some energies of the incident particle. We found that for larger numbers of electrons in the ion the effect becomes more pronounced and shifts to higher energies of the antiproton. On the other hand, a maximum of the differential cross section in the backward direction can also be found in the scattering of antiprotons on a bare uranium nucleus. The latter case can be regarded as a manifestation of the screening property of the vacuum-polarization potential
 in non-relativistic collisions of heavy particles.
\end{abstract}
\maketitle

\section{Introduction}
The FAIR facility  at GSI in Darmstadt 
 will provide high-intensity antiproton beams at energies between 30 MeV and 300 keV
 at a magnetic storage ring and at energies between 300 kev and 20 keV at an
 electrostatic storage ring. Further deceleration of antiprotons to ultra-low
 eV energies will be feasible via heavy ion trap facilities. This will enable a
 large variety of new experiments, including various atomic-collison type experiments
 with targets at storage rings. In particular,
 investigations of the antiproton scattering by a heavy ion
at low energies are anticipated with the new GSI facilities. These
investigations can give a unique possibility to observe an interesting
phenomenon predicted in Refs. \cite{gl01,gl02} and termed the Coulomb glory. 
The phenomenon consists in a prominent maximum of the differential cross section
(DCS) in  the backward direction at
a certain energy of
the incident particle, provided the interaction with a target is represented by
the Coulomb attraction of the nucleus (partly) screened by atomic electrons.
Note, that the pure Rutherford cross section shows a smooth minimum at
180$^{\circ}$.

In the present paper we examine the Coulomb glory in  collisions of antiprotons
with He-, Ne-, Ni-like, bare, and neutral uranium ($Z=92$). 
The ions are chosen to have fully occupied shells with $n=1$ (U$^{90+}$),
$n=2$ (U$^{82+}$), and $n=3$ (U$^{64+}$), respectively.
The calculations have been performed using both semiclassical and quantum theory.
Besides the screening potential due to atomic electrons, the vacuum
polarization potential is taken into account.
Atomic units ($ \hbar=e=m_e=1$) are used throughout the paper.

\section{Theoretical approach}\label{th}

We adopt the potential scattering model. This implies that neither 
excitation of atomic electrons nor ionization due to the collision 
is taken into account. The atomic electrons are regarded as a source of 
an electrostatic screening potential only. 
The use of the elastic scattering potential model can be justified for
the problem under consideration since the energy of the incident antiproton
 (which corresponds the Coulomb glory effect)
 is quite low 
and not sufficient for the excitation of core electrons in the He-, Ne-, and Ni-like
uranium ions. 
We also neglect the
polarization of the atomic electrons by the antiproton. This polarization
effect is small for highly charged uranium ions. For the neutral atom, both
polarization and inelastic processes can be significant. In the latter case,
our treatment can be considered as a first approximation retaining the most
important features of the Coulomb glory phenomenon.

Two different parts are involved in our calculations. First, we need to
construct the electrostatic potential of the target. This problem is solved
with the aid of the density-functional theory (DFT) approach. For heavy atoms such as
uranium, the relativistic version of DFT must be used. Second, we need to
calculate the scattering phaseshifts from the target potential. This part is
non-relativistic since the velocity of the antiproton is much smaller than the 
speed of light even in the vicinity of the nucleus. Moreover, the motion of the
antiproton can be treated semiclassically, and the corresponding semiclassical
methods can be used for the calculation of the phaseshifts.

\subsection{The scattering potential}
The effective potential $V(r)$ experienced by the antiproton colliding with a heavy uranium ion can be represented as a sum of three local potentials:
\begin{equation}
V(r) = V_{\rm n}(r) + V_{\rm H}(r) + V_{\rm U}(r) ,
\label{th10}
\end{equation}
where $V_{\rm n}(r)$ is the potential of an extended nucleus, $V_{\rm H}(r)$ is the Hartree potential produced by atomic electrons, and $V_{\rm U}(r)$ is the Uehling potential taking into account the effect of vacuum polarization. The potential of an extended nucleus is given by
\begin{equation}
V_{\rm n}(r) = -\int d^{3}r' \frac{\rho_{\rm n}(r')}{|\bm{r}-\bm{r}'|}.
\label{th20}
\end{equation}
Here $\rho_{\rm n}$ is the nuclear charge density, normalized by the condition
\begin{equation}
\int d^{3}r\rho_{\rm n}(r) = Z 
\label{th_nuc_norm}
\end{equation}
to the nuclear charge number $Z$.
We employ the Fermi
model
\begin{equation}
\rho_{\rm n}(r) = \dfrac{N_{0}}{1+\exp[(r-r_{0})/a]}\,,
\label{th30}
\end{equation}
where the parameter $a$ is chosen to  be $2.3\,{\rm fm}/(4 \ln3)$, 
while $r_{0}$ and  $N_0$ are derived from the root-mean-square nuclear charge radius 
and the normalization condition \cite{fricke,Shabaev02}.

The interaction of the antiproton with the atomic electrons is described by the electrostatic Hartree potential $V_{\rm H}(r)$
\begin{equation}
\begin{split}
V_{\rm H}(r) &= \int d^3 r' 
\frac{\rho(r')}
{|\bm{r}-\bm{r}'|} \\
&= \frac{4\pi}{r}\int_{0}^{r} dr'\, r'^2  \rho_{}(r') + 
4\pi\int_{r}^{\infty} dr'\, {r'}\rho_{}(r')\, ,
\end{split}
\label{th40}
\end{equation}
where $\rho(r)$ is the total electron density,
normalized via 
\begin{equation}
\int d^{3}r\rho(r) = N
\label{th55}
\end{equation}
to the total number of the atomic electrons $N$.
The electron density can be expressed in terms of the electron wave functions: 
\begin{equation}
\rho(r) \,=\,\frac{1}{4\pi} \sum_{b} \,q_{b} (g_{b}^2(r) \,+\, f_{b}^2(r)) .
\label{th60}
\end{equation}
Here $g_{b}$ and $f_{b}$ are the upper and lower radial components of the
relativistic one-electron wave functions in the shell $b$, and $q_{b}$ is the number of electrons in the shell $b$. The one-electron wave functions of heavy ions were obtained within the relativistic density functional theory using the local spin-density approximation 
with incorporation of the orbital-dependent Perdew-Zunger self-interaction correction 
\cite{Perdew}. The DFT calculations were performed employing the methods described in Refs. \cite{Tupicyn03,Tupicyn05}. 
Finally, the Uehling potential is calculated according to \cite{Shabaev02,Mohr98}
\begin{equation}
\begin{split}
&V_{U}(r)= -\frac{2\alpha^{2}}{3r}\int_{0}^{\infty}dr'
r'\rho_{n}(r')\\
&\times\int_{1}^{\infty}dt\left(1+\frac{1}{2t^{2}}\right)
\frac{\sqrt{t^{2}-1}}{t^{3}}
\\
&\times\left[\exp\left(-\dfrac{2}{\alpha}|r-r'|t\right)-
\exp\left(-\dfrac{2}{\alpha}(r+r')t\right)\right] 
\end{split}
\label{th70}
\end{equation}
with $\alpha$ being the fine structure constant.

\subsection{Calculation of the phaseshifts and differential cross sections}
 Since the expected kinetic energy of the antiproton is as low as a few hundreds of
electron volts (the maximum velocity corresponding to the classical trajectory amounts 
to about 0.01c), the non-relativistic scattering theory can be applied. 
In the present paper, we make use of the partial wave expansion of the differential cross
section
\begin{equation}
\begin{split}
\frac{d\sigma}{d\Omega}&=\frac{1}{k^{2}}\left| \frac{\nu}{2\sin^{2}\theta/2}
\exp\left(-2i\nu \ln \sin\frac{\theta}{2}\right) \right.\\
&- \sum_{l=0}^{\infty} (-1)^{l} (2l+1)
\exp(i\delta_{l}^{s})\sin\delta_{l}^{s}\\
&\left. \times\frac{(1-il/\nu)\dots (1-i/\nu)} {(1+il/\nu)\dots (1+i/\nu)}
P_{l}(\cos\theta)\right |^{2}.\label{eq1}
\end{split}
\end{equation}
Here $k$ is the momentum of the antiproton, $\nu=-Z_{c}m_{\bar{p}}/k$ is
the Coulomb parameter ($Z_{c}$ is the charge of the core and in the case of
He-like uranium $Z_{c}=90$), and $P_{l}(\cos\theta)$ are the Legendre polynomials.
As one can see, the differential cross section (DCS) is a result of the 
interference between two contributions to the total scattering amplitude: the pure
Coulomb (Rutherford) amplitude and the amplitude due to non-Coulomb
(short-range) terms in the scattering potential. The phase shifts
$\delta_{l}^{s}$ are produced by the short-range part of the scattering
potential. They can be expressed as a difference between the total phase shift
$\delta_{l}$ corresponding to the angular momentum $l$ and the Coulomb phase
shift $\delta_{l}^{c}$
\begin{eqnarray}
\delta_{l}^{s}&=&\delta_{l}-\delta_{l}^{c}\,,\\
\quad \delta_{l}^{c} &=&
\frac{1}{2i}\ln\frac{\Gamma(l+1+i\nu)}{\Gamma(l+1-i\nu)}\,.
\end{eqnarray}
A way to calculate the phase shifts $\delta_{l}^{s}$ without solving the radial
Schr\"odinger equation is provided by the variable phase method
\cite{morse33,druk49,Babikov,Babikov1,Calogero}. Under this approach, the phase shift is evaluated by
solving a first-order differential equation. The method is robust and easy to
implement. In our case, the differential equation to solve can be written as 
\begin{equation}
\begin{split}
&\frac{d}{dr}\delta_{l}^{s}(k,r)=-2m_{\bar{p}}kv(r)r^{2}\\
&\times[\cos\delta_{l}^{s}(k,r)F_{l}(k,r)
-\sin\delta_{l}^{s}(k,r)G_{l}(k,r)]^{2}\,.
\end{split}
\label{ph}
\end{equation}
Here $\delta_{l}^{s}(k,r)$ is the variable phase which depends on $r$ and $v(r)$
is the short-range part of the scattering potential
\begin{equation}
 v(r) = V(r) + \frac{Z_{c}}{r}\,.
\end{equation}
$F_{l}(k,r)$ and
$G_{l}(k,r)$ are the regular and irregular Coulomb wave functions, respectively
\cite{Abramowitz}. Eq. (\ref{ph}) is solved with the initial condition
$\delta_{l}^{s}(k,0)=0$. Then the long-distance limit of $\delta_{l}^{s}(k,r)$
gives the value of the phase shift:
\begin{equation}
\delta_{l}^{s} = \lim_{r\rightarrow\infty}\delta_{l}^{s}(k,r)\,.
\end{equation}

For the energy of the antiproton as low as a few atomic units, the absolute
value of the Coulomb parameter is large: $|\nu|\gg 1$. That means, the motion
of the antiproton can be treated in the framework of the quasiclassical
approximation \cite{Landau3}. In this limit the Coulomb wave functions in
Eq. (\ref{ph}) can be replaced by their asymptotic expressions through the
Bessel functions \cite{Abramowitz}. An alternative way to calculate the
quasiclassical phase shifts is to extract them from the quasiclassical radial
wave functions. The phase shifts $\delta_{l}^{s}$ are then expressed as a
difference of the two integrals representing the phases of the quasiclassical 
wave functions
in the total scattering potential and in the Coulomb potential $-Z_{c}/r$:
\begin{equation}
\begin{split}
\delta_{l}^{s}=\lim_{R\rightarrow\infty}\left\{\int_{r_{0}}^{R} dr
\sqrt{2m_{\bar{p}}\left(E-V(r)\right)-\frac{(l+1/2)^{2}}{r^{2}}} \right.\\
\left. -\int_{r_{c}}^{R} dr
\sqrt{2m_{\bar{p}}\left(E+\frac{Z_{c}}{r}\right)-\frac{(l+1/2)^{2}}{r^{2}}}\right\}\,,
\end{split}
\label{ph2}
\end{equation}
where $r_{0}$ and $r_{c}$ are the classical turning points for the two motions,
respectively. Note that both integrals in Eq. (\ref{ph2}) diverge as
$R\rightarrow\infty$ but their difference does not, and the phase shift is
defined correctly by the right-hand side of Eq. (\ref{ph2}). In the case $Z_{c}=0$ (neutral atom) $\delta_{l}^{s}$ are the total phase shifts corresponding to the angular momentum $l$, and Eq.~(\ref{ph2}) still can be used to determine these phaseshifts. To apply Eq.~(\ref{ph}) for the neutral atom, the Coulomb wave functions $F_{l}(k,r)$ and $G_{l}(k,r)$ have to be replaced 
by the spherical Bessel functions $j_{l}(kr)$ and $n_{l}(kr)$. 
In this case $v(r)$ represents the total (short-range) potential. 
It is also worth to
mention that the nuclear radius is much smaller than the distance
corresponding to any of the classical turning points $r_{0}$ or $r_{c}$
and,
therefore, the annihilation probability during the collision is negligible.

The quasiclassical approximation implies that the quantum number $l$ is large
\cite{Landau3}. Thus the phase shifts corresponding to small values of $l$ may
not be calculated accurately by Eq. (\ref{ph2}) (as opposed to Eq. (\ref{ph})).
However, this is not crucial since a large number of partial waves makes
comparable contributions to the scattering amplitude (in our case - hundreds
and even thousands), and the result does not depend significantly on a few
partial waves with small $l$. We have performed calculations based on both
Eqs. (\ref{ph}) and (\ref{ph2}) and found that the results agree 
with each other very well.

The more partial waves in Eq. (\ref{eq1}) contribute constructively in the
backward direction, the larger is the maximum of DCS at $\theta=180^{\circ}$.
On the other hand, the more ``classical'' is the motion of the incident
particle, the more partial waves make important presence in the wave function
and the scattering amplitude. For the same energy, this is generally the case
for heavier particles. Thus the experiments with heavy particles (antiprotons)
could be more decisive in detecting the Coulomb glory than the experiments with
light particles (electrons).

\section{Numerical results and discussion}
\begin{figure}
\vspace*{12pt}
\includegraphics[width=\columnwidth]{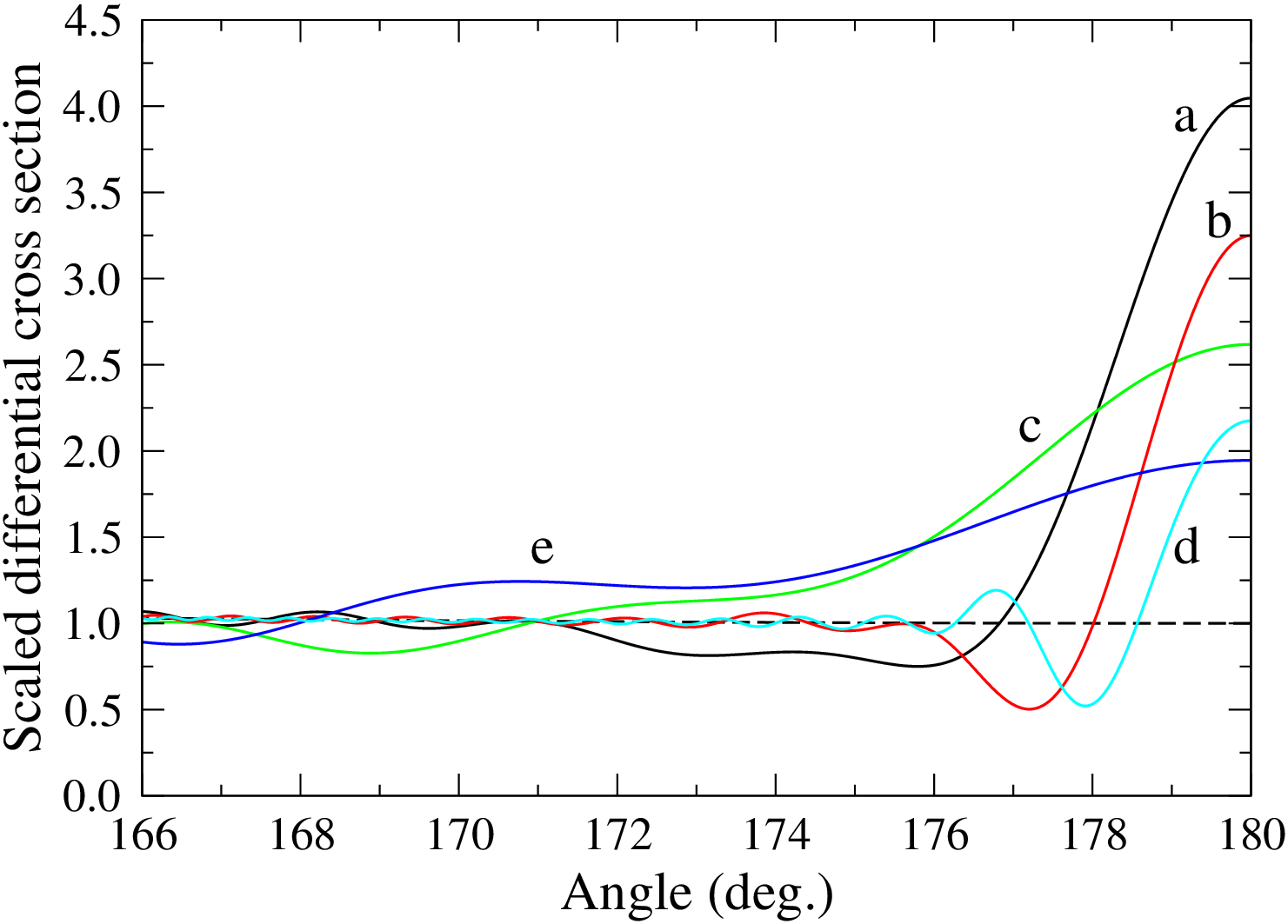}
\caption{(Color online) Scaled differential cross section $d\tilde{\sigma}/d\Omega$,
 defined by Eq. (\ref{scs}), on He-like uranium
for different energies of the incident antiproton: (a) $E=100$ eV, (b) $E=20$ eV, (c) $E=500$ eV, (d) $E=5$ eV, and (e) $E=1$ keV. The dashed line represents the scaled Rutherford cross section.}
\label{fig1}
\end{figure}

We have computed differential cross sections of antiproton--ion collisions for various energies of the antiproton and several electronic configurations of the ion. We tested both Eq.~(\ref{ph}) and Eq.~(\ref{ph2}) to calculate the phase shifts and found that they give very close results for the antiproton energies in the interval between a few tens electron volts and 10 keV, where we can expect the Coulomb glory in backward scattering. In Fig.~\ref{fig1} we present the cross sections for He-like uranium. To facilitate a comparison between the results at different energies, the differential cross sections have been scaled according to 
\begin{equation}
 \dfrac{d\tilde{\sigma}}{d\Omega} =
 \left(\dfrac{4E}{Z_{c}}\right)^{2}\dfrac{d\sigma}{d\Omega} .
\label{scs}
\end{equation}
The scaled Rutherford cross section becomes independent of energy and the ion charge:
\begin{equation}
 \dfrac{d\tilde{\sigma}^{c}}{d\Omega} =\frac{1}{\sin^{4}\theta/2}
\end{equation}
and equal to unity at $\theta=180^{\circ}$. Then the value of $d\tilde{\sigma}/d\Omega$ at $\theta=180^{\circ}$ represents the ratio of the ion DCS and the corresponding Rutherford DCS and can serve as a quantitative measure of the Coulomb glory effect. 
As one can see from Fig.~\ref{fig1}, for all the energies used in the calculations (5 eV, 20 eV, 100 eV, 500 eV, and 1 keV) $d\tilde{\sigma}/d\Omega$ as a function of the scattering angle $\theta$ has a maximum at $\theta=180^{\circ}$ that indicates the presence of the Coulomb glory. However, the strongest effect is observed at the energy $E=100$ eV when the scaled DCS reaches the value 4.1. The width of the maximum at this energy is about $2^{\circ}$. For lower and higher energies the effect becomes less pronounced. In the vicinity of the main maximum there are interference oscillations as described by Eq.~(\ref{eq1}). With increasing antiproton energy, the frequency of the oscillations becomes smaller and the main maximum broader.

\begin{figure}
\vspace*{12pt}
\includegraphics[width=\columnwidth]{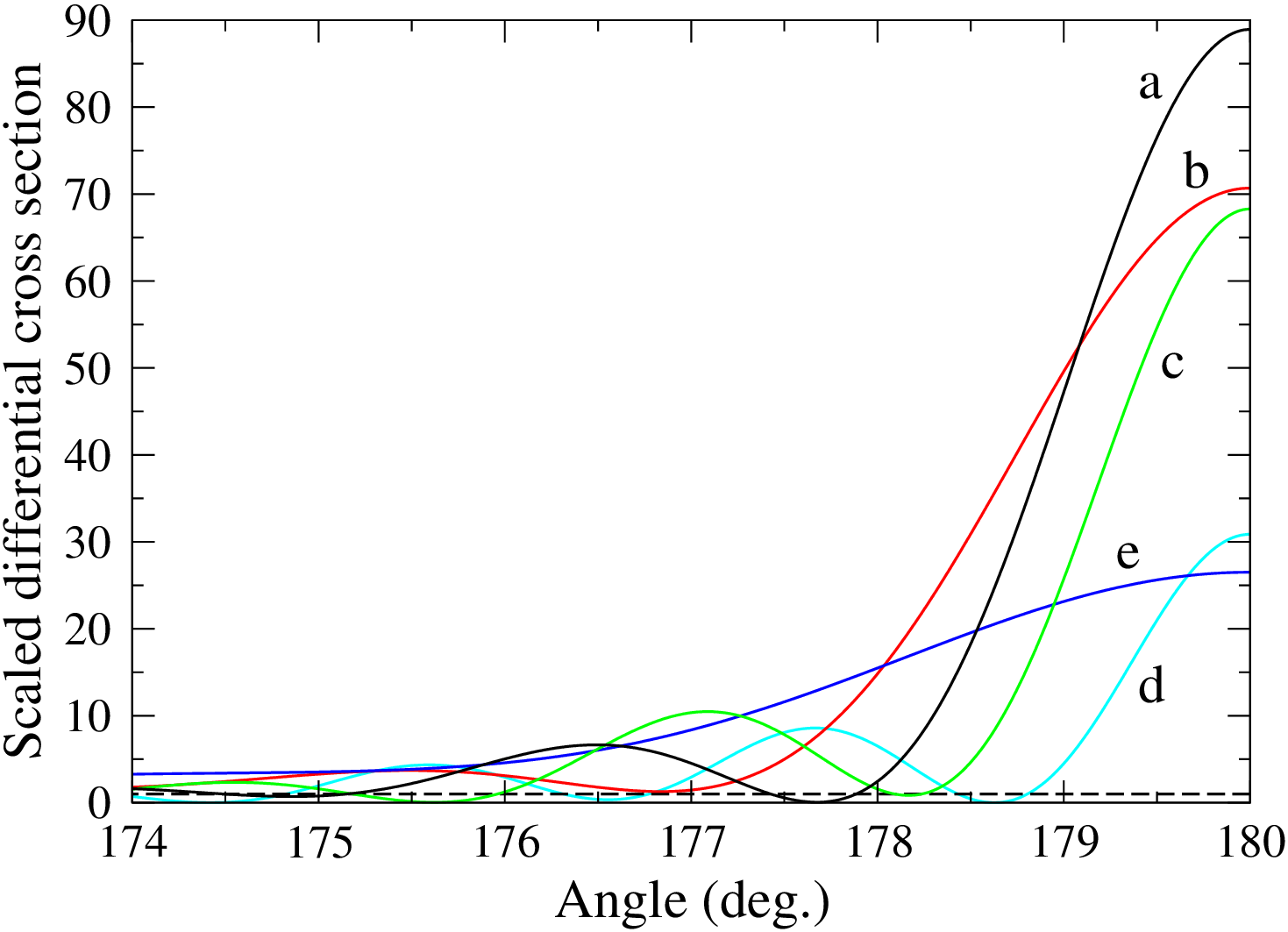}
\caption{(Color online) Scaled differential cross section $d\tilde{\sigma}/d\Omega$,
 defined by Eq. (\ref{scs}), on Ne-like uranium
for different energies of the incident antiproton: (a) $E=300$ eV, (b) $E=500$ eV, (c) $E=200$ eV, (d) $E=100$ eV, and (e) $E=1$ keV. The dashed line represents the scaled Rutherford cross section.}
\label{fig2}
\end{figure}
\begin{figure}
\vspace*{12pt}
\includegraphics[width=\columnwidth]{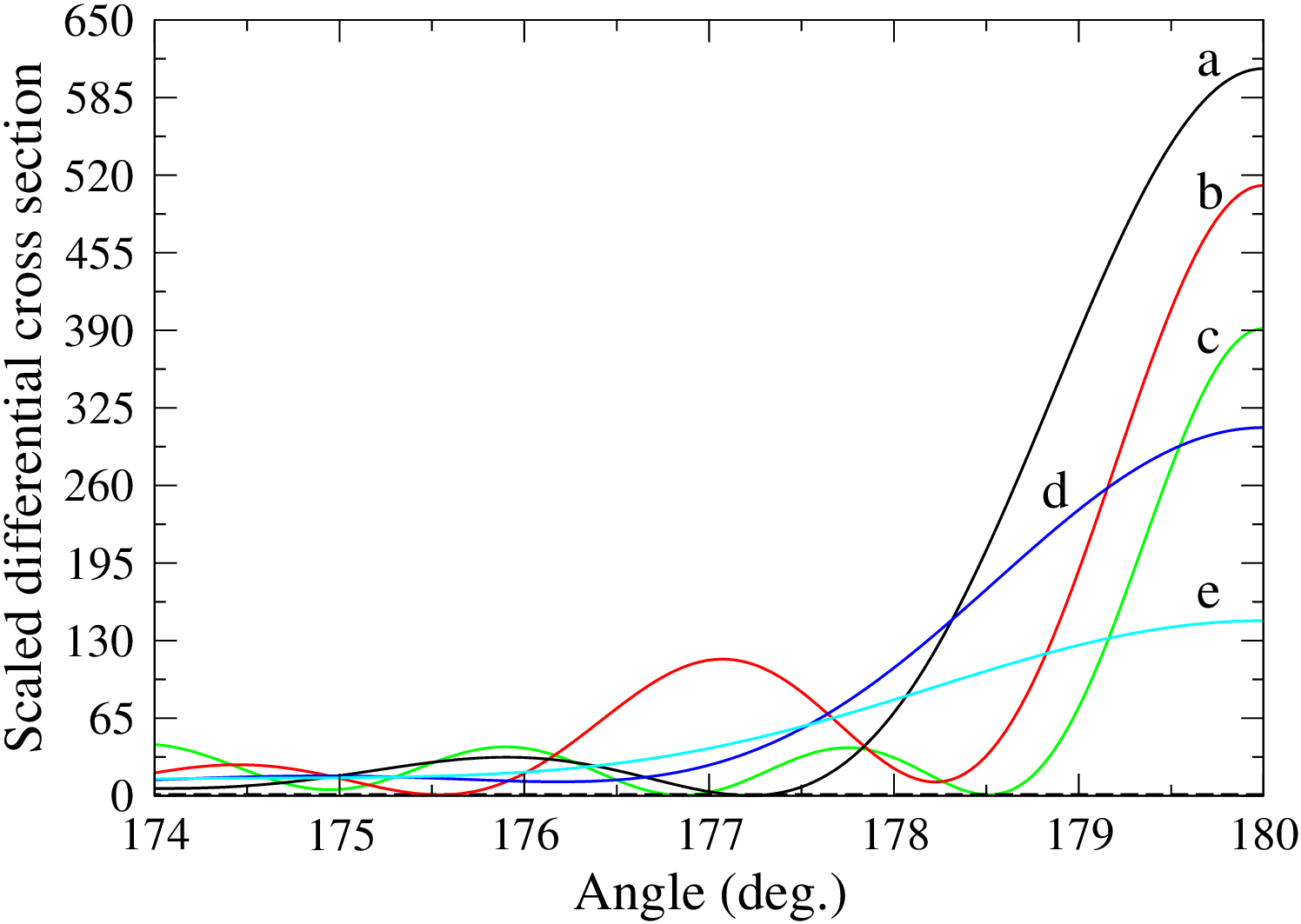}
\caption{(Color online) Scaled differential cross section 
$d\tilde{\sigma}/d\Omega$,
defined by Eq. (\ref{scs}), on Ni-like uranium
for different energies of the incident antiproton: (a) $E=2$ keV, (b) $E=1.5$ 
keV, (c) $E=1$ keV, (d) $E=2.5$ keV, and (e) $E=3$ keV. The dashed line 
represents the scaled Rutherford cross section.}
\label{fig3}
\end{figure}
The scaled DCS for Ne-like and Ni-like uranium are presented in Figs.~\ref{fig2} and \ref{fig3}, respectively. With increasing number of electrons in the ion, the range of the antiproton energies, where a prominent DCS peak exists at $\theta=180^{\circ}$, increases too. For Ne-like uranium, the Coulomb glory is best observed at the energy 300 eV with the scaled DCS equal to 90 at $\theta=180^{\circ}$. For Ni-like uranium, the corresponding  energy is 2 keV, and the scaled DCS reaches the value 609 at $\theta=180^{\circ}$. While the maximum at $\theta=180^{\circ}$ becomes higher with increasing the number of electrons and the energy of the incident antiproton, its width does not change significantly and constitutes about $1^{\circ}$.

\begin{figure}
\vspace*{15pt}
\includegraphics[width=\columnwidth]{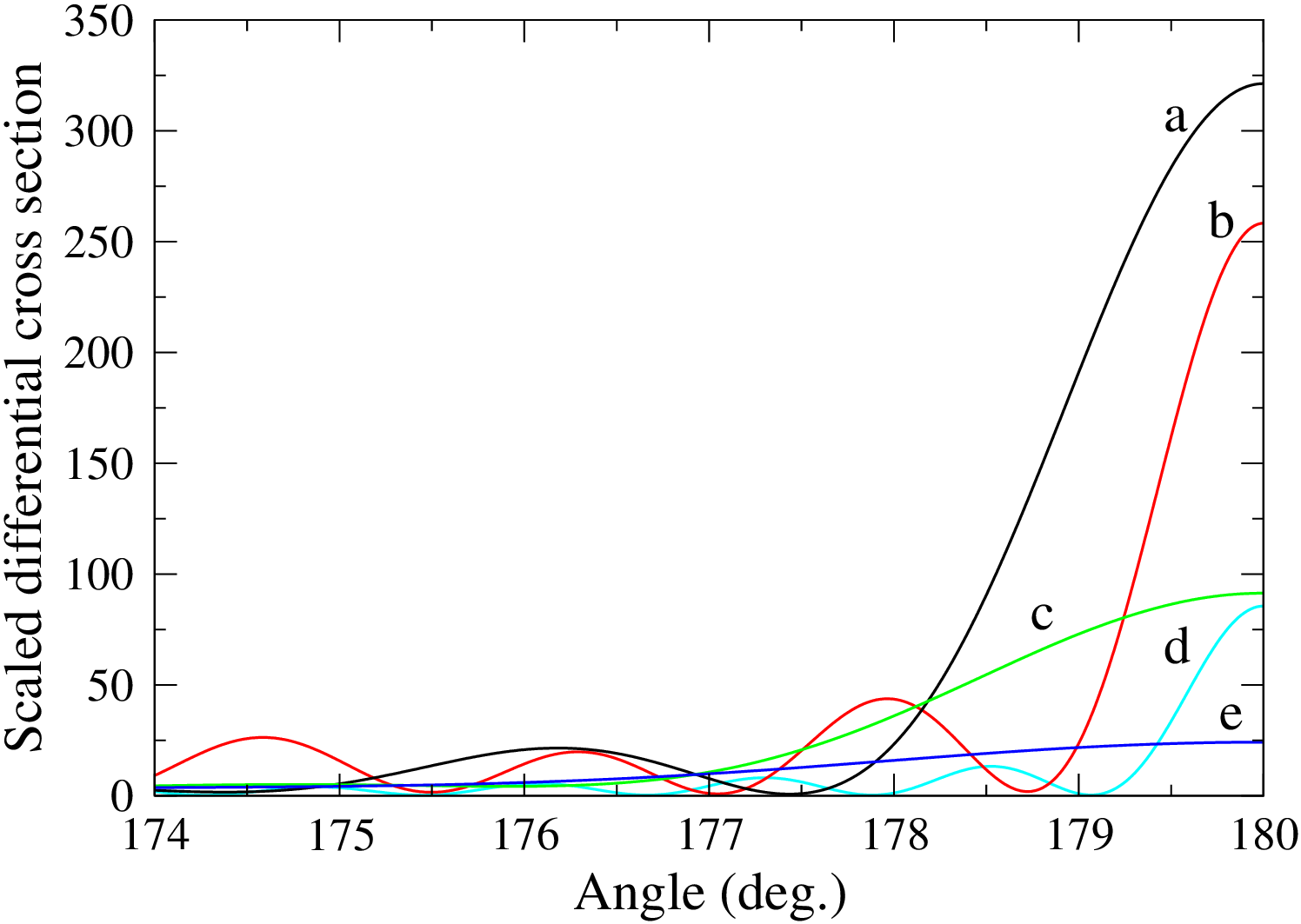}
\caption{(Color online) Scaled differential cross section $d\tilde{\sigma}/d\Omega$,
 defined by Eq. (\ref{scsn}), on neutral uranium
for different energies of the incident antiproton: (a) $E=7$ keV, (b) $E=5$ keV, (c) $E=8$ keV, (d) $E=2$ keV, and (e) $E=9$ keV.}
\label{fig4}
\end{figure}
For the uranium ions, we can compare the differential cross section with the corresponding Rutherford cross section to determine the presence of the Coulomb glory, and the quantitative expression is given by the scaled DCS (\ref{scs}). For the neutral uranium 
atom we need to find another criterion since no Rutherford cross section can be defined in this case. As a reference DCS, we use the differential cross section itself, averaged over the interval of angles between $165^{\circ}$ and $170^{\circ}$, where no prominent minima or maxima occur. 
Defined in this way, the average DCS $\langle d\sigma/d\Omega\rangle$ represents a characteristic value at a particular energy and angle range not affected by the Coulomb glory. Then the scaled DCS is defined as 
\begin{equation}
 \dfrac{d\tilde{\sigma}}{d\Omega} = \left\langle
 \dfrac{d\sigma}{d\Omega}\right\rangle^{-1} \dfrac{d\sigma}{d\Omega} \,.
\label{scsn}
\end{equation}
In Fig.~\ref{fig4} the scaled DCS is depicted  
for the neutral uranium target and several energies of the incident antiproton. The results confirm the trend already observed for uranium ions: with increasing number of electrons the Coulomb glory effect is shifted towards higher energies. For the neutral uranium, the strongest maximum at
$\theta=180^{\circ}$ corresponds to the energy 7 keV with the scaled DCS equal to 320.

\begin{figure}
\includegraphics[width=\columnwidth]{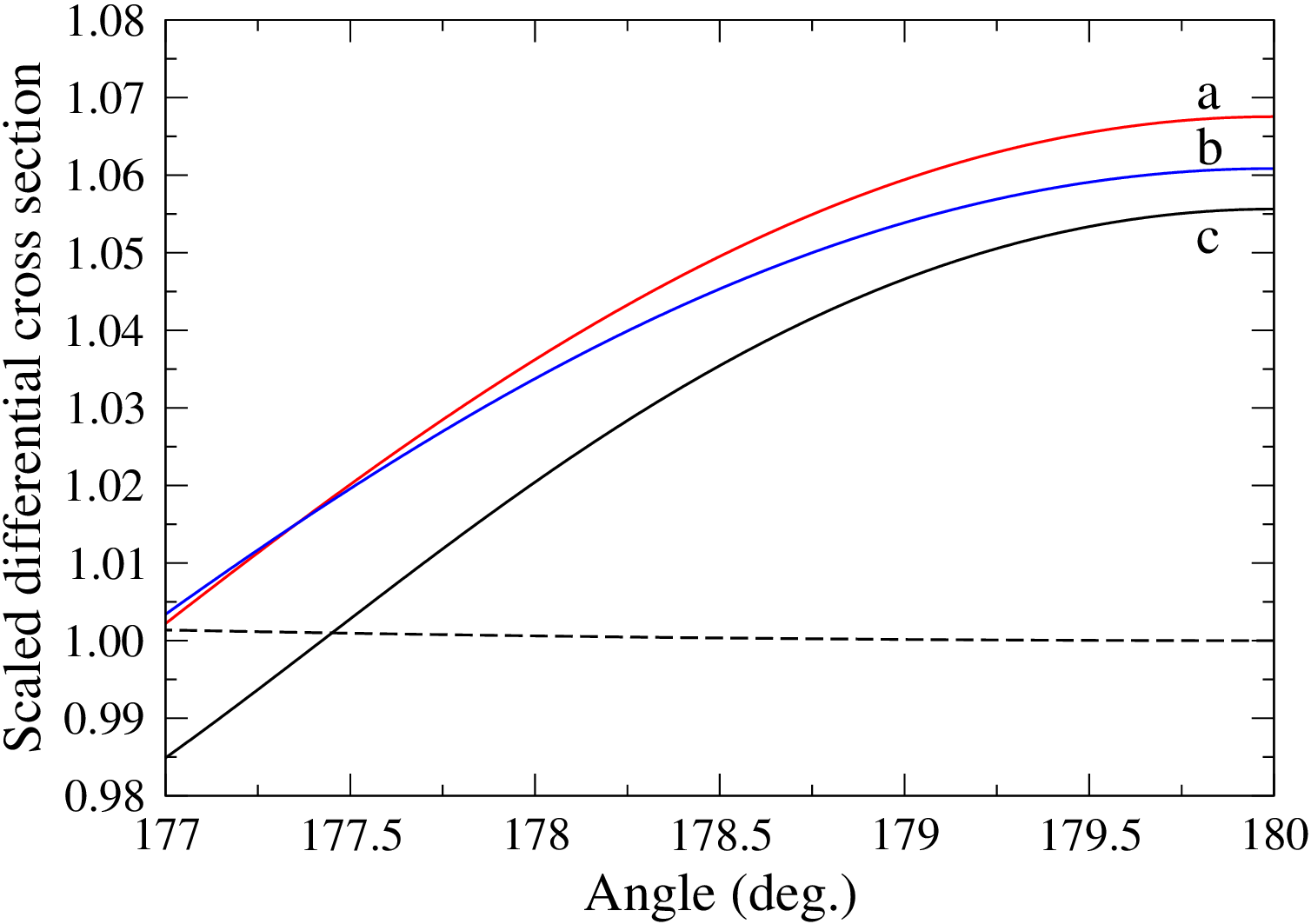}
\caption{(Color online) Scaled differential cross section $d\tilde{\sigma}/d\Omega$,
defined by Eq. (\ref{scs}), 
on bare uranium nucleus for different energies of the incident antiproton: (a) $E=400$ eV, (b) $E=800$ eV, (c) $E=100$ eV. The dashed line represents the scaled Rutherford cross section.}
\label{fig5}
\end{figure}
Observation of the Coulomb glory effect in collisions of antiprotons with bare uranium nuclei can be of particular interest. Certainly, one cannot expect a large deviation from the background Rutherford cross section because of very short range of the non-Coulomb interactions due to finite nucleus size and polarization of vacuum; the smaller the radius of the potential, the less number of phase shifts $\delta_{l}^{s}$ make noticeable contributions to the differential cross section (\ref{eq1}).  However, if such a deviation 
is detected, it becomes a direct evidence for a screening property of the vacuum-polarization potential
in non-relativistic collisions of heavy particles. Note that the finite nucleus potential is extremely short-range and influences very few first phase shifts $\delta_{l}^{s}$ only. Thus any significant deviation from the Rutherford cross section at $\theta=180^{\circ}$, which results from constructive interference of contributions with different $l$, should be mainly 
attributed to the polarization of vacuum. In Fig.~\ref{fig5} we show scaled DCS as defined by Eq.~(\ref{scs}) for the energies of the antiproton 100, 400, and 800 eV. The maximum of the scaled DCS at $\theta=180^{\circ}$ exists for all three energies while the largest deviation from the Rutherford cross section corresponds to the energy 400 eV
and amounts to about 7\%. This is, however, a rather large value, 
if we  compare its magnitude 
with a typical QED contribution to dynamical processes with heavy ions. For instance,
 the QED correction to the DCS for the radiative recombination of an electron by a bare uranium
 nucleus does not exceed the level of about 2\% \cite{Shabaev02,shabaev00}. 

\section{Conclusion}
In this work we have studied the backward scattering  of low-energy antiprotons by highly charged and neutral uranium. We found that a maximum in the differential cross section at the scattering angle $\theta=180^{\circ}$ exists in a wide range of energies of the incident particle. However, at some energies the effect is enhanced. Classical mechanics describes this phenomenon as a combination of glory and rainbow scattering at particular energies; it was termed the Coulomb glory \cite{gl01,gl02}. Our quantum-mechanical calculations showed that the Coulomb glory can be observed for the energies of the antiproton within the range 100 eV -- 7 keV, depending on the electronic configuration of the ion. In general holds, the larger the number of electrons, the higher the energy where the effect has its strongest manifestation. In the case of 
Coulomb glory, the differential cross section at $\theta=180^{\circ}$ can be much larger than the corresponding background cross section. Actually, the ratio of these two quantities ranges from 4 for the He-like uranium to 609 for the Ni-like uranium. 
We also investigated possible manifestation of the effect in collisions of antiprotons with bare uranium nuclei. 
In this case the scattering potential differs from the Coulomb potential due to finite-nuclear size   and vacuum-polarization effects. Both interactions are of very short-range that prevents large DCS values in the backward direction. However, some deviation from the Rutherford cross section does exist and can be increased by appropriate tuning of the antiproton energy. If experimentally detected, this effect can be regarded as an interesting manifestation of the screening character of the vacuum-polarization potential. 

\acknowledgments
We thank Prof. Yu. N. Demkov for drawing our attention to the Coulomb glory effect
with heavy ions and for stimulating conversations.
Valuable discussions with Dr. W. Quint are also acknowledged.
This work was partially supported by INTAS-GSI (Grant No. 06-1000012-8881),
RFBR (Grant No. 07-02-00126a), GSI, and DFG.
A.V.M. also acknowledges the support from  DAAD and the Dynasty foundation.

\end{document}